\documentclass[aps,pra,twocolumn,showpacs,amsmath,amssymb,floatfix,superscriptaddress,notitlepage]{revtex4-1}

\usepackage{color}
\usepackage{bbm}
\usepackage{graphicx}% Include figure files
\usepackage{dcolumn}% Align table columns on decimal point
\usepackage{bm}% bold math
\usepackage{array}
\usepackage{float}
\usepackage{supertabular}
\usepackage{longtable}
\usepackage{mathrsfs}
\usepackage{txfonts}
\usepackage{wasysym}
\usepackage[T1]{fontenc}
\usepackage[latin1]{inputenc}
\usepackage{amssymb}
\usepackage[usenames,dvipsnames]{xcolor}
\usepackage{amsmath}
\usepackage{cases}
\usepackage{amstext}
\usepackage{latexsym}
\usepackage[colorlinks=true,citecolor=Cerulean,linkcolor=RubineRed,urlcolor=Cerulean]{hyperref}
\usepackage{enumitem}

\begin{document}
\title{
 Nonexponential quantum decay under environmental decoherence
}

\author{M. Beau}
\affiliation{Department of Physics, University of Massachusetts, Boston, MA 02125, USA}
\affiliation{Dublin Institute for Advanced Studies, School of Theoretic
al Physics, 10 Burlington Road, Dublin 4, Ireland}
\author{J. Kiukas}
\affiliation{Department of Mathematics, Aberystwyth University, Aberystwyth SY23 3BZ, UK}
\author{I. L. Egusquiza}
\affiliation{Department of Theoretical Physics and History of Science,\\ University of the Basque Country UPV/EHU, Apartado 644, 48080 Bilbao, Spain}
\author{A. del Campo}
\affiliation{Department of Physics, University of Massachusetts, Boston, MA 02125, USA}

%%%%%%%%%%%% Equations %%%%%%%%%%%%%%%
\newcommand{\beq}{\begin{equation}}
\newcommand{\eeq}{\end{equation}}
\newcommand{\beqa}{\begin{eqnarray}}
\newcommand{\eeqa}{\end{eqnarray}}

%%%%%%%%%%%% Maths notations %%%%%%%%%%%%%%%
\newcommand{\intf}{\int_{-\infty}^\infty}
\newcommand{\into}{\int_0^\infty}
\def\tr{{\rm Tr}}
\def\S{\mathcal{S}}
\def\E{\mathcal{E}}
\newcommand{\bmat}{\begin{pmatrix}}
\newcommand{\emat}{\end{pmatrix}}

%%%%%%%%%%%% Brackets %%%%%%%%%%%%%%%
\def\l{\left}
\def\r{\right}
\def\la{\langle}
\def\ra{\rangle}
\def\da{\dagger}

%%%%%%%%%%%% Scripts %%%%%%%%%%%%%%%

\begin{abstract}
A  system prepared in an unstable quantum state generally decays following an exponential law, as environmental decoherence is expected to prevent the decay products from recombining to reconstruct the initial state. Here we show the existence of deviations from exponential decay in open quantum systems  under very general conditions.  Our results are illustrated with the exact dynamics under quantum Brownian motion and suggest an explanation of recent experimental observations.
\end{abstract}
\maketitle

The exponential decay law of unstable systems is ubiquitous in Nature and  has widespread applications \cite{Gamow,WW30,Adler03}. Yet, in isolated quantum systems deviations occur at both short and long times of evolution \cite{FGR78,Schulman08,Gorin06}.
Short time deviations underlie the quantum Zeno effect \cite{MS77,shortexp}, ubiquitously used to engineer decoherence free-subspaces and preserve quantum information.
Long-time deviations are expected in any  non-relativistic systems with a ground state; they slow down the decay and  generally manifest as a power-law  in time \cite{Khalfin57}. Both short and long-time deviations are present as well in many-particle systems \cite{delcampo11,Goold11,PSD12,Longhi12,IA12,delcampo16}.
Indeed, the latter  signal the advent  of thermalization in isolated many-body systems \cite{TS14,TS15}. In quantum cosmology, power-law deviations constrain the likelihood of scenarios with eternal inflation \cite{KD08}.
 They also rule the scrambling of information  as measured by the decay of the form factor \cite{PR15,Cotler16,Dyer16,delcampo17} in blackhole physics and strongly coupled quantum systems described by AdS/CFT, that are believed to be maximally chaotic \cite{Maldacena15}.

Given a unstable quantum state $|\Psi_0\ra$ prepared at time $t=0$, it is customary to describe the closed-system decay dynamics via the survival probability, which is the fidelity between the initial state and its time evolution
\beqa
\label{surv}
\mathcal{S}(t):=|\mathcal{A}(t)|^2=|\la\Psi_0|\Psi(t)\ra|^2.
\eeqa
Explicitly, the survival amplitude reads $\mathcal{A}(t)=\la\Psi_0|\hat{U}(t,0)|\Psi_0\ra$, where $\hat{U}(t,0)=T\exp(-i\int_0^tds\hat{H}(s)/\hbar)$ is the time evolution operator generated by the Hamiltonian of the system $\hat{H}$.
Short time deviations are associated with the quadratic decay 
\beqa
\mathcal{S}(t)=1-(t/\tau_Z)^2+\mathcal{O}(t^3), 
\eeqa
and are generally suppressed by the coupling to an environment that induces the appearance of a  term linear in $t$, see, e.g. \cite{WW30,Adler03,Halliwell14,Chenu17}.
The origin of the long-time deviations can be appreciated using the Ersak equation for the survival amplitude \cite{Ersak69,FG72,FGR78}
\beqa
\label{ersakeq}
\mathcal{A}(t)=\mathcal{A}(t-t')\mathcal{A}(t')+m(t,t'),
 \eeqa
 that follows from the unitarity of time evolution in isolated quantum systems.
The memory term  reads
 \beqa
 \label{memun}
 m(t,t')=\la \Psi_0|U(t,t')\hat{Q}U(t',0)|\Psi_0\ra,
\eeqa
Here, we denote the projector onto the space spanned by the initial state by $\hat{P}\equiv|\Psi_0\ra\la\Psi_0|$ and its orthogonal complement by $\hat{Q}\equiv 1-\hat{P}$.
As a result, the memory term $m(t,t')$ represents the formation of decay products at an intermediate time $t'$ and their subsequent recombination to reconstruct the initial state $|\Psi_0\ra$.
The suppression of this term leads to the exponential decay law for $\mathcal{A}(t)$ and $\mathcal{S}(t)$, as an ansatz of the form $\mathcal{A}(t)=e^{-\gamma t}$  is a solution of Eq. (\ref{ersakeq}) with $m(t,t')=0$, i.e, $\mathcal{A}(t)=\mathcal{A}(t-t') \mathcal{A}(t')$. \cite{FG72,FGR78}. 
In addition, using the definition of  the survival probability and Eq. (\ref{ersakeq}), it has been demonstrated that the long-time non-exponential behavior of  $\mathcal{S}(t)$ is dominated by $|m(t,t')|^2$. 
The onset of long-time deviations generally occurs after many lifetimes, making their direct observation  challenging. This has motivated the quest for systems where the decay is dominated by deviations and exponential decay is absent, see e.g., \cite{GCV06}.  Nonexponential decay actually governs the dynamics in the absence of resonant states, e.g.,  under free dispersion.  

The breakdown of unitarity can lead to exponential behavior for arbitrarily long times, as it happens in non-Hermitian systems with complex energy eigenvalues \cite{Gamow,complexV, KRSW12}. Nonhermitian Hamiltonians can be justified when the dynamics is restricted to a given subspace as well as in quantum measurement theory, and can delay or suppress nonexponential decay \cite{Muga06}. More generally, environmentally-induced decoherence is widely believed to suppress quantum state reconstruction and deviations from the exponential law \cite{FGR78,BF98}, as shown in  quantum optical systems \cite{Knight76,KM76}, see as well \cite{BF17}. In view of this, it came as a surprise that experimental observations consistent with nonexponential decay were reported  in an open quantum system \cite{Monkman06}.

In this work, we show that nonexponential decay is ubiquitous in open quantum systems, i.e., even in the presence of environmental decoherence. We show that state reconstruction is to be expected under Markovian dynamics and is responsible for the breakdown of the exponential law. While the short-time behavior is consistent with exponential quantum decay, long-time deviations subsequently occur. These deviations are explicitly  illustrated 
in the decay under  quantum Brownian motion.

%Whenever the environment monitors the decay products, it can enhance state reconstruction via quantum Zeno effect. 

{\it Markovian dynamics.---}
Consider the Hilbert space $\mathcal{H}_{SE}=\mathcal{H}_S\otimes\mathcal{H}_E$ obtained via the tensor product of the Hilbert space for the system $\mathcal{H}_S$ and that of  the environment $\mathcal{H}_E$.  The dynamics in $\mathcal{H}_{SE}$ is described by a unitary time evolution operator $\hat{U}_{SE}(t,0)$ generated by the full Hamiltonian, 
$\hat{H}_{SE}=\hat{H}_{S}+\hat{H}_{E}+\hat{H}_{int}$,
where $\hat{H}_{int}$ denotes the interaction between the system and the environment.
The evolution of an initially factorised state $\rho_{SE}(0)=\rho_S(0)\otimes\rho_E$ is described by the von Neumann-Liouville equation
\beqa
\rho_{SE}(t)=\hat{U}_{SE}(t,0)\rho_S(0)\otimes\rho_E\hat{U}_{SE}(t,0)^\dag
\eeqa
from which the reduced density matrix of the system $\rho_S(t)=\tr_E\rho(t)$ is obtained by 
tracing over the environmental degrees of freedom. Under weak coupling, the evolution of the reduced dynamics of the system, $\rho_S(t)=V(t)\rho_S(0)$, is Markovian in the sense that $V(t)$ is a quantum dynamical semigroup with the composition property $V(t)V(t')=V(t+t')$ for $t,t'\geq 0$. For  clarity of presentation, we shall focus on the case where $\rho_S(0)$ is pure and refer to the Supplemental material \cite{SMqbrown} for the mixed  case with $\tr\rho_S(0)^2<1$.

{\it Short-time Markovian asymptotics.---} Given a dynamical semigroup $V(t)$, the master equation associated with it  is of  Lindblad form \cite{Lindblad76,Breuer02}
\beq\label{LindbladMasterEq}
\frac{d}{dt}\rho_S = \frac{-i}{\hbar}\l[\hat{H}_s,\rho_S\r]+ \sum_\alpha \gamma_\alpha \left[L_\alpha\rho_S L_\alpha^\dag-\frac{1}{2}\left\{L_\alpha^\dag L_\alpha,\rho_S\right\}\right]\ ,
\eeq
where $L_\alpha$ are the Lindblad operators. Consider the fidelity between the initial pure state $\rho_S(0)=|\Psi_0\ra\la\Psi_0|$ and the time dependent state $\rho_S(t)=V(t)\rho_S(0)$,
\beqa
\label{SPeq}
 \mathcal{S}(t):=F[\rho_S(0),\rho_S(t)]=\la \Psi_0|\rho_S(t)|\Psi_0\ra.
 \eeqa
Explicit computation shows that the exact short-time asymptotics is given by \cite{SMqbrown}
\beqa
\mathcal{S}(t)= 1-\frac{t}{\tau_D}+\mathcal{O}(t^2) ,
\eeqa
where 
\beqa
\label{tauD}
\tau_D=\frac{1}{\sum_{\alpha}\gamma_\alpha {\rm Cov}\left(L_\alpha,L_\alpha^\dag\right)},
\eeqa
and the covariance of two operators $A$ and $B$ is defined as $ {\rm Cov}(A,B)=\la AB\ra-\la A\ra\la B\ra$.   The requirement that the initial state is pure can be lifted, see \cite{SMqbrown}.  
This universal behavior of the short-time dynamics for Markovian open quantum systems is consistent with an exponential decay and suggests the identification of $\tau_D^{-1}$ with the decay rate.

By contrast, the short time asymptotic of the  survival probability in $\mathcal{H}_{SE}$, defined as $\mathcal{S}_{SE}(t):=F[\rho_{SE}(0),\rho_{SE}(t)]=\left[\tr\sqrt{\rho_{SE}(0)^\frac 12\rho_{SE}(t)\rho_{SE}(0)^\frac 12}\right]^2$,   is   characterized under  Hamiltonian dynamics by a sub-exponential decay
\beqa
\mathcal{S}_{SE}(t) = 1- F_0 t^2/4+\mathcal{O}(t^3), 
\eeqa
where the positive constant $F_0>0$ is the quantum Fisher information $F_0={\rm tr}[\rho_{SE}(0) \mathrm{L}_0^2]$ defined via the symmetric logarithmic derivative $\mathrm{L}_t$, that satisfies $\frac{d}{dt}\rho_{SE}(t) = (\mathrm{L}_t\rho_{SE}(t)+\rho_{SE}(t)  \mathrm{L}_t)/2$ \cite{Hubner93,Paris09}.  The subexponential decay of $\mathcal{S}_{SE}(t)$ has important applications and can be exploited, e.g. to slow down or accelerate the decay  \cite{Sakurai,KK00}.
While it is known that the Markovian master equation fails generally at very short-times,  
within the realm of its validity short-time deviations are  absent. In what follows, we shall focus on  the nonexponential behavior in the subsequent dynamics, under Eq. (\ref{LindbladMasterEq}).

{\it Quantum state reconstruction under quantum dynamical semigroups.---}
Using the composition property of dynamical semigroups it is possible to derive an analogue of the Ersak equation (\ref{ersakeq}) for open quantum systems. 
This generalization requires a formulation  in terms of probabilities, simplifying the interpretation of the analogue of the memory term in the unitary case, (\ref{memun}).
Indeed, explicit computation yields
\beqa
 \mathcal{S}(t)&=&\tr[\hat{P}V(t)\rho_S(0)]\\
&=&\tr[\hat{P}V(t-t')(\hat{P}+\hat{Q})V(t')\rho_S(0)(\hat{P}+\hat{Q})]\\
&=& \mathcal{S}(t-t') \mathcal{S}(t')+ M(t,t'),\label{GenErsak}
\eeqa
where we have used the fact that $\hat{P}\rho_S(t')\hat{P}=S(t')\hat{P}$ and introduced the memory term
\beqa
\label{Mterm}
M(t,t')&=&\tr\left\{\hat{P}V(t-t')\left[\hat{Q}\left(V(t')\rho_S(0)\right)\hat{Q}\right]\right\}\nonumber\\
& & +\tr\left\{\hat{P}V(t-t')\left[\hat{Q}\left(V(t')\rho_S(0)\right)\hat{P}\right]\right\}\nonumber\\
& & +\tr\left\{\hat{P}V(t-t')\left[\hat{P}\left(V(t')\rho_S(0)\right)\hat{Q}\right]\right\}.
\eeqa
Equation  (\ref{GenErsak}) is the generalization of the Ersak equation  \cite{Ersak69} for quantum Markovian dynamics for an initial pure state $\rho_S(0)=\hat{P}$; see \cite{SMqbrown} for the mixed case. The first term in the rhs of the memory term $M(t,t')$  (\ref{Mterm}) represents the conditional probability to find the time-evolving state at time $t$ in the space spanned by the initial state, provided it was found in the orthogonal subspace at an intermediate time $t'$, i.e., that it had fully decayed. The remaining two crossed terms result from interference involving state reconstruction, i.e., the coherences of the density matrix for this $P$-$Q$ decomposition. 
When the memory term vanishes identically, $M(t,t')=0$, the generalized Ersak  equation (\ref{GenErsak}) dictates  exponential decay,  $ \mathcal{S}(t)=e^{-\gamma t}$, that satisfies $ \mathcal{S}(t)=\mathcal{S}(t-t') \mathcal{S}(t')$. As in the unitary case \cite{FGR78}, any deviation from an exponential decay law for $ \mathcal{S}(t)$ arises due to the state reconstruction of the initial state $\rho_S(0)$  from the decay products found at the intermediate time $t'$, during the evolution between $t'$ and $t$. 
The memory term does not generally vanish, justifying the ubiquity of deviations from the exponential decay law in open quantum systems.

{\it Universality of long-time subexponential decay in Markovian quantum systems.---}
We next establish that the long-time decay of the survival probability in open  quantum systems, in particular also Markovian, is generally not exponential. We focus on Hamiltonians $\hat{H}_{SE}$ with a continuous energy spectrum $E\in  [E_0,\infty)$. In general, each energy eigenvalue may have several (improper) eigenstates associated with it, but this multiplicity does not play any role in the following, and hence we assume for simplicity that there is just one eigenstate $|E\rangle$ for each $E$. One can easily check that the same argument works also in the general case.

An initial state of the composite system of factorized form $\rho_{SE}(0)=\rho_S(0)\otimes\rho_E$ will have coherences in the energy representation. We write it in its diagonal representation $\rho_{SE}(0)=\sum_j\lambda_j|\lambda_j\ra\la\lambda_j|$ where the occupation numbers $\lambda_j\geq 0$ and $ |\lambda_j\ra \in\mathcal{H}_{SE}$.
We next exploit a purification of $\rho_{SE}(0)$ in an enlarged Hilbert space $\mathcal{H}_{SE}\otimes\mathcal{H}_{R}$; we take $\mathcal H_R=\mathcal{H}_{SE}$ and define
 \beqa
 |\Psi_{SER}(0)\ra=\sum_j\sqrt{\lambda_j}|\lambda_j\ra\otimes |\lambda_j\ra,
\eeqa
where the finiteness of the sum $\sum_j \lambda_j = {\rm tr}[\rho_{SE}(0)]=1<\infty$ ensures that this belongs to the Hilbert space. 
Denoting by $|E\ra\in \mathcal{H}_{SE}$ the energy eigenkets of $\hat{H}_{SE}$, we consider the time evolution operator
\beqa
\hat{U}_{SE}(t,0)\otimes 1_R=\int_{E_0}^{\infty}dE e^{-iEt/\hbar}|E\ra\la E|\otimes  1_R.
\eeqa
The dynamics of the purified state is then described by
 \beqa
 |\Psi_{SER}(t)\ra &=&(\hat{U}_{SE}(t,0)\otimes 1_R)|\Psi_{SER}(0)\ra\nonumber\\
 &=&\sum_j\int_{E_0}^{\infty}dE\sqrt{\lambda_j}\la E|\lambda_j\ra e^{-iEt/\hbar}|E\ra\otimes |\lambda_j\ra,
\eeqa
and we stress that $\Psi_{SER}(t)$ is indeed a purification of the physical system-environment state $\rho_{SE}(t)= \hat{U}_{SE}(t,0)\rho(0)\hat{U}_{SE}(t,0)^\dagger$ (and hence also of the system state $\rho_S(t)$) at each time $t$. In terms of $\Psi_{SER}(t)$ we introduce the survival amplitude between the initial purified state at $t=0$ and that at any time $t\geq0$, i.e., 
\beqa
\mathcal{A}_{SER}(t)&:=&\la\Psi_{SER}(0)|\Psi_{SER}(t)\ra\nonumber\\
&=&\sum_j\int_{E_0}^{\infty}dE\lambda_j|\la E|\lambda_j\ra|^2e^{-iEt/\hbar}\nonumber\\&=& \int_{-\infty}^\infty dE\,\varrho_{SE}(E)\,e^{-iEt/\hbar},
\eeqa
where we have denoted the energy distribution of the initial state by $\varrho_{SE}(E)=|\la\Psi_{SE}(0)|E\ra|^2=\sum_j\lambda_j|\la E|\lambda_j\ra|^2$, a function that vanishes  for any $E<E_0$. We note that the combination of the sum and the integral converges absolutely, hence the order may be interchanged. 
%Two important consequences follow. First, following Fock and Krylov \cite{FK47}, we noticed that $\widetilde{\varrho_{SE}}(E)=\varrho_{SE}(E)\Theta(E-E_0)$ is an $L^1$-function, i.e., $\int  \widetilde{\varrho_{SE}}(E)dE<\infty$. As a result, its Fourier transform  vanishes identically at long times of evolution; $\mathcal{A}_{SER}(t)\rightarrow 0$ as $t\rightarrow\infty$. 
%Therefore, open quantum systems described by a Hamiltonian $\hat{H}_{SE}$ with a continuous energy spectrum $E\in  [E_0,\infty)$ lead to full decay in the sense that $|\mathcal{A}_{SER}(t)|^2$ vanishes asymptotically.
%Second, 
The semi-finiteness of $\varrho_{SE}(E)$ dictates the analytic properties of its Fourier transform $\mathcal{A}_{SER}(t)$. In particular, the Paley-Wiener theorem \cite{PW34,Khalfin57}  imposes the convergence of the integral
\beqa
\int_{\mathbb{R}} dt\frac{|\log|\mathcal{A}_{SER}(t)||}{1+(t/t_0)^2}<\infty\ ,
\eeqa
where  $t_0$ is  any  constant with dimensions of time.
As a result, the survival probability in the enlarged Hilbert space decays slower than any exponential function $e^{-\alpha t}$ for large $t$, 
\beq\label{decay}
|\mathcal{A}_{SER}(t)|^2\geq Ce^{-\gamma t^q}, \ \text{ with } C,\gamma>0, \text{  and } q<\ ,
\eeq
 In order to connect this with the survival probability $\mathcal{S}(t)$ we use Uhlmann's theorem, which states that the fidelity $\mathcal{S}_{SE}(t):=F[\rho_{SE}(0),\rho_{SE}(t)]$ equals the maximal fidelity between all possible purifications of $\rho_{SE}(0)$ and $\rho_{SE}(t)$ \cite{Josza94, Uhlmann76}. Since $\Psi_{SER}(t)$ is a purification of $\rho_{SE}(t)$, we get $\mathcal{S}_{SE} \geq |\mathcal A_{SER}(t)|^2$ for each $t\geq 0$. 
Finally, using the monotonicity (or non-contractivity) of the fidelity, $\mathcal{S}(t)\geq \mathcal{S}_{SE}$, it follows that the the survival probability of the system $\mathcal{S}(t)$ cannot decay faster than the fidelity in $\mathcal{H}_{SE}$. We note that the theorem \cite{Uhlmann76} is not restricted to a finite-dimensional setting, and hence works in our case. From \eqref{decay} we then get
\beq\label{decay2}
\mathcal{S}(t)\geq Ce^{-\gamma t^q}, \ \text{ with } C,\gamma>0, \text{  and } q<1.
\eeq
which is the main result of this section and generalizes the corresponding result for closed systems \cite{FGR78}. 
To summarize, the existence of the ground-state  $E_0$  in the composite system  $\hat{H}_{SE}$ makes $\varrho_{SE}(E)$ semi-finite and dictates the long-time behavior of the survival probability $\mathcal{S}(t)$ of the system $\rho_S(0)$ under Markovian evolution.  While the result holds in full generality, it is intended for the case of an infinite-dimensional system Hilbert space $\mathcal H_S$. In fact, in the finite-dimensional case $\mathcal S(t)$ is not even expected to vanish at long times, since $\rho_S(t)$ typically tends to a full-rank stationary state. Only quantities such as coherences can then exhibit exponential decay (as in \cite{BF17}).

We have established that deviations from exponential decay are to be expected under environmental decoherence at both short and long-times of evolution. In particular, a  power-law behavior as  experimentally reported in \cite{Monkman06} is consistent with open quantum dynamics. Yet, a nearly exponential decay is not excluded by the Paley-Wiener theorem \cite{Iwo98}.
We next focus on a paradigmatic example of open quantum dynamics on an infinite-dimensional space and that is governed by nonexponential behavior: quantum Brownian motion.

{\it Nonexponential decay under quantum  Brownian motion.---}
Consider a single quantum particle of mass $m$ in contact with a thermal bath. We assume weak coupling between the particle and the bath, a large temperature regime, and the Born-Markov approximation. The dynamics is then well described by the Caldeira-Leggett model \cite{Breuer02,Weiss08} 
\beq\label{MasterEq}
\frac{d}{dt}\rho_S(t)=\frac{-i}{\hbar}\l[ H,\rho_S(t)\r]-\frac{i\gamma}{\hbar} [x,\{p,\rho_S(t)\}]-D\l[ x,\l[x,\rho_S(t)\r]\r]\ ,
\eeq
with 
$
H=-\frac{\hbar^2}{2m}\frac{\partial^2}{\partial x^2} \ ,
$
and where the coupling constant $D=2m\gamma k_B T/\hbar^2$ depends explicitly on the temperature $T$ of the bath and on the damping constant $\gamma$.  Eq. (\ref{tauD}) for the decoherence time  leads to
\beq\label{DecohTimeQBM}
\tau_D = \frac{\lambda_\beta^2}{2\gamma{\Delta} x^2}\ ,
\eeq
when  $\tau_D\ll\gamma^{-1}$, in terms of the de Broglie thermal wavelengths $\lambda_\beta^2 = \hbar^2/(2m k_B T)$ and \( \Delta x^2 \)  the variance of the initial pure state.
Consistently with the high temperature regime, we assume that the characteristic time scale of the system $\tau_c\equiv m\Delta x^2/\hbar$ is large compared to the thermal bath characteristic time $\tau_\beta \equiv \hbar/(k_B T)$. Equivalently, $\lambda_\beta \ll \Delta x$ which entails  $\tau_D\ll\tau_R= \gamma^{-1}$.

For the sake of illustration,  consider the initial pure Gaussian state 
\beq\label{InitialGaussian}
\rho_S(x,y;0) = \sqrt{\frac{1}{\pi\sigma^2}} \exp{\l[-\l( \frac{ x^2+ y^2}{2\sigma^2} \r)\r]} \ ,
\eeq
with $\Delta x=\sigma/\sqrt{2}$. 
 We find  that there are three distinct regimes in the decay dynamics:
 
(i) At short times $ t \ll \tau_D \ll \tau_R$,   the survival probability  behaves as $S(t)\approx 1-t/\tau_D$.

(ii)  Subsequently, an intermediate regime sets in for $ \tau_D \ll t \ll \tau_R$ when the system is undamped and experiences decoherence. The off-diagonal density matrix decays exponentially in time \cite{Zurek91} 
$
\rho_S(x,y;t)\approx \rho_S(0,0;t)\times \exp{\l[-D t (x-y)^2\r]} \exp{\l[i\phi(x,y)\r]} \ ,
$
where $\phi$ is the complex part of the phase. 
In this regime, the density profile (this is, the diagonal part of $\rho_S(x,y;t)$) has the asymptotics 
$
\rho_S(x,x;t) \approx \frac{1}{\sqrt{2\pi \Delta x(t)^2}} \exp{\l(-\frac{x^2}{2\Delta x(t)^2}\r)}\ ,
$
where the normalization factor of the density matrix scales as \cite{Breuer02}
$
\Delta x(t) \approx \sqrt{\frac{4Dt^3}{3}}\ .
$
Hence, we find  from the long-time asymptotics of density matrix  $\rho_S(x,y;t)\approx\sqrt{\frac{2}{Dt}}\delta(x-y)\rho_S(x,x;t)$ that
the survival probability \eqref{SPeq} simplifies to $\mathcal{S}(t)=\intf dx\intf dy \rho_S(x,y;0)\rho_S(y,x;t)\approx \sqrt{\frac{2}{Dt}}\intf dx \rho_S(x,x;0)\rho_S(x,x;t) \approx \sqrt{\frac{2}{Dt}}\frac{1}{\Delta x(t)}$ as $\Delta x(t)^2\gg \Delta x$. This leads to the power-law asymptotic behavior
\beq\label{S(t)Approx2LT1}
\mathcal{S}(t) \approx \sqrt{3}\frac{m}{D\hbar t^2}=\sqrt{3}\frac{\tau_\beta \tau_R}{t^2} \ ,
\eeq
which is independent of   ${\Delta} x$ and the initial width of the Gaussian wave packet. The two relevant time scales are set by $\tau_\beta \equiv \hbar/(k_B T)$ and $\tau_R=\gamma^{-1}$. 
%Experimentally, one would measure the overlap between the initial and long-time density of states. 
%Therefore we use the unnormalized survival probability defined by equation \eqref{SPeq}. Notice that $\mathcal{S}(t)$ could be normalized by dividing equation \eqref{SPeq} by the purity $p_0$.

(iii) The final and third stage of the dynamics occurs when  the evolution time becomes large compared to the relaxation time of the environment $t\gg \tau_R\gg\tau_D$. The system becomes overdamped and the off-diagonal density matrix converges to a stationary solution
$
\rho_S(x,y;t)\approx \rho_S(0,0;t)\times \exp{\l(-\frac{(x-y)^2}{2\lambda_\beta^2}\r)} \exp{\l[i\phi(x,y)\r]}\ ,
$
where the normalization factor is given by the diffusion variance 
$
\Delta x(t) \approx \sqrt{\frac{D\hbar^2 }{m\gamma^2}t} =\sqrt{\frac{2k_B T}{m\gamma} t}\ ,
$
see \cite{Breuer02}. Using similar arguments to those for (ii) we find the long-time power-law scaling 
\beq\label{S(t)Approx2LT2}
\mathcal{S}(t) \approx \frac{\lambda_\beta}{\Delta x(t)}\approx \frac{\hbar\gamma}{2k_B T}\frac{1}{\sqrt{\gamma t}} = \frac{\tau_\beta}{\tau_R}\sqrt{\frac{\tau_R}{t}}\ .
\eeq
Equations \eqref{S(t)Approx2LT1}-\eqref{S(t)Approx2LT2} can also be derived from an asymptotics  analysis of the exact $\mathcal{S}(t)$, found via the Feynman-Vernon influence functional \cite{FV63,Breuer02,Weiss08}  and the multi-dimensional Gaussian integral method; see \cite{SMqbrown}. 
As anticipated, the asymptotic behavior predicted  by equations \eqref{S(t)Approx2LT1}-\eqref{S(t)Approx2LT2} does not depend on the initial spreading $\Delta x$. A direct application of our findings is the estimation of the damping coefficient $\gamma$ and the temperature $T$ of the system, that can be extracted from  the survival probability \eqref{SPeq} upon identification of the two time scales $\tau_\beta=\hbar/(k_B T)$ and $\tau_R=\gamma^{-1}$. To do this it suffices to find the intercepts of the two asymptotic lines obtained in a log-log plot, see Fig \ref{Fig1} and  equations \eqref{S(t)Approx2LT1}-\eqref{S(t)Approx2LT2}.

\begin{figure}
 \includegraphics[width= 0.9\columnwidth]{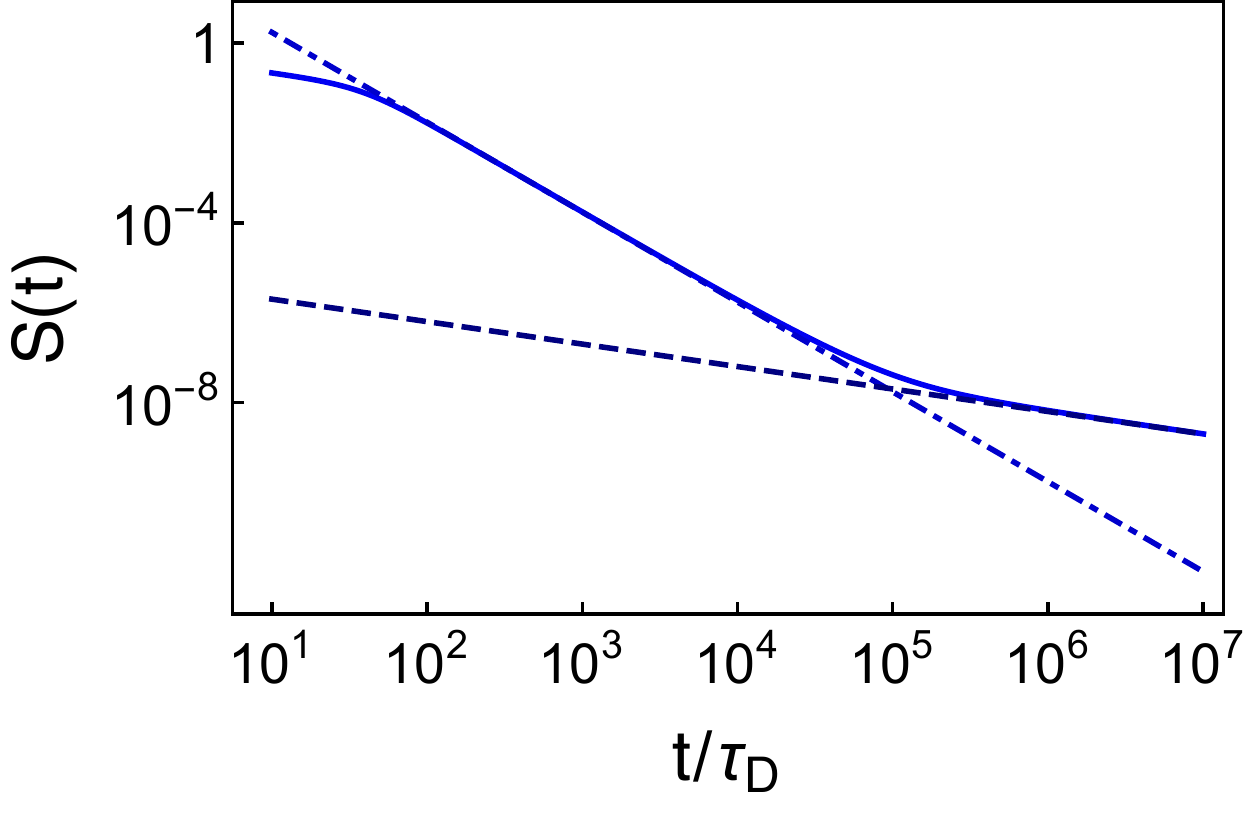}\ \ \ \\
\caption{{\bf Decay of the survival probability under quantum Brownian motion.}  The change in the power-law governing the survival probability of a pure initial Gaussian state (solid line) is shown in a log-log scale as a function of time in units of the decoherence time.  The long-time asymptotic expressions for $t\gg\tau_D$ in the two distinct regimes $\gamma t\ll 1$ (dotted-dashed line) and $\gamma t\gg 1$ (dashed line) are also shown, with $\gamma = 10^{-3}$ and $D\sigma^2=100$, with  $\hbar/m\sigma^2$ as a unit of frequency.  }\label{Fig1}
\end{figure}

We next consider another prominent example -the quantum decay of a Schr\"odinger cat state-  that supports the idea of universality of the long-time asymptotic behavior \eqref{S(t)Approx2LT1}-\eqref{S(t)Approx2LT2}. Consider a pure state $\rho_S(x,y;0)=\psi_0(x)\psi_0(y)^\ast$ made of a superposition of two Gaussian wave packets centered respectively at $x=-r$ and $x=+r$, i.e., 
\beq\label{TwoGaussianState}
\psi_0(x) = \mathcal{N}_\sigma \left[e^{-\frac{(x-r)^2}{2\sigma^2}} + e^{-\frac{(x+r)^2}{2\sigma^2}} \right],
\eeq
where the normalization factor is
$
\mathcal{N}_\sigma = \left[2\sqrt{\pi\sigma^2}(1+e^{-\frac{r^2}{\sigma^2}})\right]^{-1/2}\ .
$
The arguments we used  for the Gaussian state \eqref{InitialGaussian} apply as well to the cat state \eqref{TwoGaussianState}. Inserting the expression for the  variance of the position 
$
\Delta x^2 = \sigma^2/2+r^2/\left(1+e^{-\frac{r^2}{\sigma^2}}\right)\ 
$
into equation \eqref{tauD}, we reproduce the  prediction by Zurek, $\tau_D=\lambda_\beta^2/(2\gamma r^2)$, in the limit $r\gg \sigma$ \cite{Zurek91}. In the  limit $r\ll \sigma$, we find $\tau_D=\lambda_\beta^2/(\gamma \sigma^2)=1/(D\sigma^2)$ which agrees with the  Bedingham-Halliwell decoherence time derived from the short-time asymptotics in  \cite{Halliwell14}. The intermediate cases $r\sim\sigma$ offer new regimes  for any initial state with a finite variance $\Delta x$. While the decoherence time defines the long-time scaling, it does not appear in the expression of the asymptotic survival probability  \eqref{S(t)Approx2LT1}-\eqref{S(t)Approx2LT2}. This remarkable result combined with our previous findings for Gaussian states of the form \eqref{InitialGaussian} supports the idea of universality of long-time asymptotics, that should be experimentally testable. The change in the power-law scaling of $\mathcal{S}(t)\propto 1/t^2$ to  $\mathcal{S}(t)\propto 1/\sqrt{t}$  is demonstrated in in Fig \ref{Fig1} for for $r=0$. We  obtain similar plots for $r>0$, that indicate the universal behavior of the long-time quantum decay of  the survival probability. 

{\it Summary.---}
We have shown that the  decay dynamics of  open quantum systems generally exhibits deviations from exponential decay in the presence of environmental decoherence.  These deviations result from the reconstruction of the initial state from the decay products formed during the course of the evolution. 
While the short-time quantum decay under Markovian dynamics is consistent with an exponential law, the long-time evolution is characterized by a sub-exponential decay whenever a ground state exists for the system-environment  complex, e.g., in a nonrelativistic setting. We have demonstrated the existence of  these deviations  in quantum Brownian motion, a setting amenable to experimental investigations. Our study is expected to find broad applications across a wide variety of fields. Prominent instances include the analysis of  decoherence dynamics, collapse models in quantum measurement theory,  thermalization and information scrambling in  open quantum systems, and  quantum cosmology.

{\it Acknowledgments.---} 
Funding support from UMass Boston (project P20150000029279), ESF (POLATOM-5052) and the John Templeton Foundation is  acknowledged. I.L.E. acknowledges funding from Spanish MINECO/FEDER Grant No. FIS2015-69983-P, and Basque Government IT986-16. AdC acknowledges the hospitality of the Centre for Quantum Technologies at the  National University of Singapore during the completion of this work.

\newpage

\appendix

\begin{widetext}

\tableofcontents

\section{Markovian master equation and short-time quantum decay}

We consider the Lindblad form of  the Markovian master equation describing the dynamics of a quantum open system \cite{Lindblad76,Breuer02}
\beq\label{MasterEq}
\frac{d}{dt}\rho_S(t) = \frac{-i}{\hbar}\l[H,\rho_S(t)\r]+ \sum_\alpha\gamma_\alpha\l(L_\alpha \rho_S(t) L_\alpha^\dagger - \frac{1}{2}L_\alpha^\dagger L_\alpha\rho_S(t)- \frac{1}{2}\rho_S(t) L_\alpha^\dagger L_\alpha\r)\ ,
\eeq
where $L_\alpha$ are the Lindblad operator.

We define the relative purity $S(t)$ between the initial and time-dependent reduced density matrices $\rho_S(0)$ and $\rho_S(t)$ as an analogue of the survival probability,
\beq\label{S(t)}
S(t) \equiv \frac{\text{tr}\Big[\rho_S(0)\rho_S(t)\Big]}{\text{tr}\Big[\rho_S(0)^2\Big]} \ ,
\eeq
where the trace in the denominator represents the purity of the initial state which is equal to $1$ if the state is initially pure, i.e., when the relative purity becomes identical to the survival probability.
Let us denote the unnormalized relative purity by $\tilde{S}(t)=\text{tr}\Big[\rho_S(0)^2\Big]S(t)$. Its first derivative reads
\beq
\frac{d}{dt}\tilde{S}(t) = \frac{-i}{\hbar}\text{tr}\Big\{[H,\rho_S(t)]\rho_S(0)\Big\} + \sum_\alpha\gamma_\alpha\l\{
\text{tr}\Big[L_\alpha\rho_S(t) L_\alpha^\dagger\rho_S(0)\Big]-\frac{1}{2}\text{tr}\Big[L_\alpha^\dagger L_\alpha\rho_S(t)\rho_S(0)\Big]-\frac{1}{2}\text{tr}\Big[L_\alpha^\dagger L_\alpha\rho_S(0)\rho_S(t)\Big]\r\}\ ,
\eeq
with the limit
\beqa\label{DerSurvivalt0}
\tilde{S}'(0)&\equiv& \lim_{t\rightarrow 0}\frac{d}{dt}\tilde{S}(t) =  - \sum_\alpha\gamma_\alpha\l\{\text{tr}\Big[L_\alpha^\dagger L_\alpha\rho_S(0)^2\Big]-
\text{tr}\Big[L_\alpha\rho_S(0) L_\alpha^\dagger\rho_S(0)\Big]\r\} = - \sum_\alpha\gamma_\alpha\widetilde{\text{Cov}}_{\rho_S(0)}\l( L_\alpha^\dagger , L_\alpha\r)\ ,
\eeqa
where the modified covariance is defined as 
$
\widetilde{\text{Cov}}_{\rho_S(0)}\l(A,B\r) \equiv \langle AB\rho_S(0) \rangle - \langle A \rho_S(0) B \rangle\ ,
$
and $ \langle X \rangle \equiv \text{tr}\l[X\rho_S(0)\r] ,\ X=A,B$. In equation \eqref{DerSurvivalt0} we have $A=  L_\alpha$ and $B=L_\alpha^\dagger$. 
As a result, the  short-time asymptotic expansion reads
\beq\label{S(t)order1}
S(t)= 1-\frac{t}{\tau_D}+\mathcal{O}(t^2)\ ,
\eeq
with 
\beq\label{tauD}
\tau_D=\frac{\text{tr}\l[\rho_S(0)^2\r]}{\sum_{\alpha}\gamma_\alpha\widetilde{\text{Cov}}_{\rho_S(0)}\l( L_\alpha^\dagger , L_\alpha\r)}\ .
\eeq
The previous equations shows universal behavior of short-time dynamics for Markovian open quantum systems and generalize the result in \cite{CBCdC16} to non-Hermitian Lindblad operators and initial mixed states. In turn, this result will prove key to obtain a long-time asymptotic for the quantum Brownian motion. 

Let us consider different relevant cases:
\begin{itemize}
\item For Hermitian Lindblad operators $L_\alpha=L_\alpha^\dagger$ we can introduce the modified variance denoted $\widetilde{\Delta} L_\alpha^2$ 
\beq\label{VarModif}
\widetilde{\Delta} L_\alpha^2 \equiv \widetilde{\text{Cov}}_{\rho_S(0)}\l( L_\alpha , L_\alpha\r) = \text{tr}\Big[L_\alpha^2\rho_S(0)^2\Big]-\text{tr}\Big[\l(L_\alpha\rho_S(0) \r)^2\Big]\ ,
\eeq
which is equal to the variance of the operator
$$
\Delta L_\alpha^2\equiv \text{tr}\Big[L_\alpha^2\rho_S(0)\Big]-\text{tr}\Big[L_\alpha\rho_S(0) \Big]^2 = \la L_\alpha^2 \ra - \la L_\alpha\ra^2\ ,
$$
when $\rho_S(0)$ is pure, in  agreement with the result derived in \cite{CBCdC16}.
\item For the general case where $L_\alpha$ can be non-Hermitian, if $\rho_S(0)$ is pure we find
\beq
\widetilde{\text{Cov}}_{\rho_S(0)}\l( L_\alpha^\dagger,L_\alpha\r) =\text{Cov}_{\rho_S(0)}\l( L_\alpha^\dagger,L_\alpha\r) =  \langle L_\alpha^\dagger L_\alpha \rangle - \langle L_\alpha^\dagger \rangle \langle L_\alpha \rangle  \ ,
\eeq
where the standard covariance is
$
\text{Cov}_{\rho_S(0)}\l(A,B\r) \equiv \langle AB\rangle - \langle A \rangle \langle B \rangle\ . 
$
\end{itemize}

\section{Exact short-time quantum decay of  a system-environment composite state}

The   survival probability in the system-environment Hilbert space $\mathcal{H}_{SE}$ is defined as $\mathcal{S}_{SE}(t):=F[\rho_{SE}(0),\rho_{SE}(t)]$. As we next show, it is generally characterized by a subexponential decay. When the initial system-environment state $\rho_{SE}(0)$ is pure this result follows from elementary quantum dynamics \cite{Sakurai}. Under the unitary evolution generated by the full Hermitian Hamiltonian $\hat{H}_{SE}=\hat{H}_{S}+\hat{H}_{E}+\hat{H}_{int}$, the time evolution operator can be expanded to second order in time, and an explicit computation yields
\beqa
\label{savar}
\mathcal{S}_{SE}(t) = 1- \frac{\Delta H_{SE}^2}{\hbar^2} t^2+\mathcal{O}(t^3),
\eeqa
where  $\Delta H_{SE}^2:=\langle H_{SE}^2\rangle -\langle H_{SE}\rangle^2$ is the variance of the Hamiltonian $H_{SE}$ in the initial pure state $\rho_{SE}(0)$.

More generally, the initial state $\rho_{SE}(0)=\rho_{S}(0)\otimes\rho_{E}(0)$ is expected to be mixed, as $\rho_{E}(0)$  describes an environment.
To derive the short-time asymptotics of $\mathcal{S}_{SE}(t)$ we  note that for any two states, 
\beqa
\mathcal{S}_{SE}(t):=F[\rho_{SE}(0),\rho_{SE}(t)]=\left(1-\frac 12 D_B[\rho_{SE}(0),\rho_{SE}(t)]^2\right)^2,
\eeqa
where $D_B$ is the Bures distance \cite{Hubner93}. For two neighbouring states (i.e. for small $t$), we can use the differential form 
\beqa
D_B[\rho_{SE}(0),\rho_{SE}(t)]^2 = F_0 t^2/4 + O(t^3),
\eeqa
 where 
$F_0$ is the quantum Fisher information of $\rho_{SE}(t)$ at $t=0$ \cite{Hubner93}. Hence we get
\beqa
\mathcal{S}_{SE}(t) = 1- F_0 t^2/4+\mathcal{O}(t^3).
\eeqa
 The quantum Fisher information is given by $F_0={\rm tr}[\rho_{SE}(0) \mathrm{L}_0^2]$ where the symmetric logarithmic derivative $\mathrm{L}_t$ is determined by the differential equation $\frac{d}{dt}\rho_{SE}(t) = (\mathrm{L}_t\rho_{SE}(t)+\rho_{SE}(t)  \mathrm{L}_t)/2$, and can be explicitly computed (see the above references). When $\rho_{SE}(0)$ is pure, we get
\beqa
F_0 = 4\Delta H_{SE}^2/\hbar^2,
\eeqa
 in agreement with the well-known result for a pure initial state, Eq. (\ref{savar}). In the case where $\rho_{SE}(0)$ is mixed, the quantum Fisher information is no longer related to $\Delta H_{SE}^2$; see e.g. \cite{Paris09}.

\section{Generalized Ersak equation for Markovian semigroups}

The concept of projector for pure states has an extension to mixed states using the Hilbert--Schmidt (HS) inner product. Namely, given a reference state \(\rho\) (which is HS), any HS operator \(A\) admits a decomposition of the form
\begin{equation}
  \label{eq:bdecomp}
  A= \frac{\mathrm{Tr}\left( \rho A\right)}{\mathrm{Tr}\left(\rho^2\right)} \rho + A^{\perp}\,,
\end{equation}
such that
\begin{equation}
  \label{eq:aperp}
  \mathrm{Tr}\left( \rho A^{\perp}\right)=0\,.
\end{equation}
In other words, any state  \(\rho\) gives rise to a projector in the HS space,
\begin{equation}
  \label{eq:projector}
  \mathcal{P}A= \frac{ \mathrm{Tr}\left( \rho A\right)}{ \mathrm{Tr}\left( \rho^2\right)} \rho\,.
\end{equation}
It is easy to check that it is indeed a projector, \(\mathcal{P}^2=\mathcal{P}\) and positivity.

Let us now assume that the evolution of the initial reduced density matrix of the system $\rho_S(0)$  is governed by a dynamical semigroup \begin{equation}
  \label{eq:semigroup}
  \rho_S(t)=V(t)\rho_S(0)\,,
\end{equation}
with \(V(t+t')=V(t)V(t')\). Then the standard proof for the Ersak equation goes through for the relative purity (\(\mathcal{P}_0\) is the projector in HS associated with \(\rho_S(0)\)), here introduced as
\begin{equation}
  \label{eq:purity}
  S(t) = \mathrm{Tr}\left[ \mathcal{P}_0 \rho_S(t)\right]= \frac{ \mathrm{Tr}\left[ \rho_S(0)\rho_S(t)\right]}{ \mathrm{Tr}\left[ \rho_S(0)^2\right]}\,.
\end{equation}
Namely, the proof goes as follows
\begin{eqnarray*}
  S(t)&=& \mathrm{Tr}\left[\mathcal{P}_0 V(t) \rho_S(0)\right]\ \\
  &=& \mathrm{Tr}\left[\mathcal{P}_0 V(t-t')\left(\mathcal{P}_0+\mathcal{Q}_0\right)V(t') \rho_S(0)\right]\ \\
      &=& \mathrm{Tr}\left\{\mathcal{P}_0 V(t-t')\left[ \mathcal{P}_0V(t')\rho_S(0)\right]\right\}+ M(t,t')\ \\
  &=& \mathrm{Tr}\left[\mathcal{P}_0 V(t-t') \rho_S(0)\right] \frac{ \mathrm{Tr}\left[\rho_S(0)V(t')\rho_S(0)\right]}{\mathrm{Tr}\left[\rho_S(0)^2\right]}+M(t,t')\ \\
  &=& S(t-t')S(t')+ M(t,t')\,.
\end{eqnarray*}
Clearly \(\mathcal{Q}_0\) is the projector complementary to \(\mathcal{P}_0\), i.e. \(\mathcal{P}_0+\mathcal{Q}_0=1\).

The structure of the memory term is analogous to the case for pure states: %, and can again be interpreted as the conditional probability of reconstruction given that it went through a (HS) orthogonal state at intermediate time \(t'\).
\begin{equation}
  \label{eq:memoryterm}
  M(t,t')= \mathrm{Tr}\left\{\mathcal{P}_0 V(t-t')\left[ \mathcal{Q}_0V(t')\rho_S(0)\right]\right\}\,.
\end{equation}

Notice that the expansion of the memory term in the main text as contributions coming from different reconstruction paths is not available here. In the main text the derivation was carried out for pure states, and the projectors \(P\) and \(Q\) act on the Hilbert space; on the other hand, the HS projectors \(\mathcal{P}_0\) and   \(\mathcal{Q}_0\) are superoperators acting on operators. The derivation presented here is much more closely connected to the standard derivation of the Ersak equation for \emph{amplitudes}.

\section{Quantum Brownian motion and non-exponential decay}
 
\subsection{Caldeira-Leggett model}

Consider a single particle of mass $m$ in a time-dependent harmonic trap with frequency $\omega(t)$ in contact with a thermal bath. Let us further assume  weak coupling between the particle and the bath, a large temperature regime, and the Born-Markov approximation. Under these conditions,  the Caldeira-Leggett master equation governs the dynamics of the reduced density matrix   $\rho_S(t)$ \cite{Breuer02}, 
\beq\label{MasterEq}
\frac{d}{dt}\rho_S(t)=\frac{-i}{\hbar}\l[ H,\rho_S(t)\r]-\frac{i\gamma}{\hbar} [x,\{p,\rho_S(t)\}]-D\l[ x,\l[x,\rho_S(t)\r]\r]\ ,
\eeq
with 
\beq\label{QHO}
H=-\frac{\hbar^2}{2m}\frac{\partial^2}{\partial x^2} + \frac{m\omega(t)^2}{2}x^2\ ,
\eeq
and where the coupling constant $D=2m\gamma k_B T/\hbar^2$ depends explicitly on the temperature $T$ of the bath and on the damping constant $\gamma$.  It will be convenient at points  to write it in the form \( D= \gamma/ \lambda^2_\beta \), with \( \lambda_\beta=\hbar/\sqrt{2m k_BT} \) the thermal de Broglie wavelength.
In the coordinate representation, the master equation \eqref{MasterEq} reads
\beq\label{MasterEq2}
\frac{\partial}{\partial t}\rho_S(x,y;t)=\frac{i\hbar}{2m}\l(\partial_x^2-\partial_{y}^2\r)\rho_S(x,y;t)-\frac{im\omega(t)^2}{2\hbar}\l(x^2-y^2\r)\rho_S(x,y;t)-\gamma (x-y)\l(\partial_x-\partial_{y}\r)\rho_S(x,y;t)-D\l(x-y\r)^2\rho_S(x,y;t)  \ ,
\eeq
where $\rho_S(x,y;t)\equiv \la x|\rho_S(t)|y\ra$.

To solve the dynamics exactly, we use the path integral method to compute the Green's function of the density matrix in the position representation, i.e., of the so-called the Feynman-Vernon influence functional \cite{FV63,Breuer02}. It reads
\beq\label{J}
J\l(x,y;x_0,y_0;t\r) = J_{0}(x,y;x_0,y_0;t) e^{\Gamma(x,y;x_0,y_0;t)}\ ,
\eeq
where $J_{0}(x,y;x_0,y_0;t)$ is the damped evolution (complex phase)
\beqa
J_{0}(x,y;x_0,y_0;t) &=& \frac{m}{2\pi\hbar G_2(t)}\\
& & \times\exp\l\{\frac{im}{2\hbar G_2(t)}\l[\l(x-x_0\r)\l(\dot{G}_2 x-x_0\r)-\l(y-y_0\r)\l(\dot{G}_2 y-y_0\r)+\l(xx_0 - yy_0\r)\l(1+g(t)\r)+\l(xy_0-yx_0\r)\l(1-g(t)\r)\r]\r\}\nonumber
\eeqa
and the function $g(t)  = 1-\epsilon +\epsilon \dot{G}_2(t)$ with $\epsilon = \left\{ \begin{array}{ll} 0 \ \text{for}\ \gamma=0 \\  1 \ \text{for}\ \gamma\neq 0 \end{array} \right.$. This means that cross terms $xy_0$ and $yx_0$ disappear when there is no damping $\gamma=0$, which is expected. 
The real phase (decoherence) is
\beqa
\Gamma(x,y;x_0,y_0;t) = -D \l[ \alpha(t)(x_0-y_0)^2 + \beta(t)(x-y)^2+2\eta(t)(x-y)(x_0-y_0) \r]\ , 
\eeqa
where 
\begin{subequations}\label{functionsJ}
\beq
\alpha(t) = \int_{0}^t ds \frac{G_2(t-s)^2}{G_2(t)^2}\ ,
\eeq
\beq
\beta(t) = \frac{1}{G_2(t)^2}\int_{0}^t ds \l[G_2(t)G_1(t-s)-G_1(t)G_2(t-s)\r]^2\ ,
\eeq
\beq
\eta(t) = \frac{1}{G_2(t)}\int_{0}^t ds \l[G_2(t-s)G_1(t-s)-\frac{G_1(t)}{G_2(t)}G_2(t-s)^2\r]  \ ,
\eeq
\end{subequations}
and where the functions $G_1(t)$ and $G_2(t)$ are the fundamental solutions of the classical equation
\beq\label{ClassEq}
\ddot{x}(t)+2\gamma \dot{x}(t)+\omega(t)^2 x(t) = 0\ ,
\eeq
fulfilling the following boundary conditions 
\begin{gather*}
G_1(0)=1,\ \dot{G}_1(0)=0 \ , \\
G_2(0)=0,\ \dot{G}_2(0)=1 \ ,
\end{gather*}
and the Wronskian identity
$$
W(G_1(t),G_2(t)) \equiv \dot{G}_1(t)G_2(t) - \dot{G}_2(t)G_1(t) = \left\{ \begin{array}{ll} -1 \ \text{for}\ \gamma=0 \\  -\dot{G}_2(t) \ \text{for}\ \gamma\neq 0 \end{array} \right.\ . 
$$

The solution of the Master equation \eqref{MasterEq2} is obtained via the identity
\beq\label{rhoSol}
\rho_S(x,y;t)=\intf dx_0\intf dy_0 J\l(x,y;x_0,y_0;t\r)\rho_S(x_0,y_0;0)\ .
\eeq

\subsection{Scaling dynamics of a Gaussian state and exact evolution of the density matrix}

For $D=0$ (no decoherence), the dynamics satisfies scale invariance. Given an initial pure state $\rho_S(0) = |\psi_n\ra\la\psi_n|$ where $|\psi_n\ra$ is the $n$-eigenstate of $H$,  the matrix element of the density matrix reads 
\beq\label{ScalingD=0}
\rho_S(x,y;t)=  \frac{1}{b(t)^{1/2}}e^{i\frac{m\dot{b}(t)}{2\hbar b(t)}(x^2-y^2)}\psi_n\l(\frac{x}{b(t)}\r)\psi_n^\ast\l(\frac{y}{b(t)}\r)\ ,
\eeq 
where the scaling factor $b(t)$ is the unique solution of the Ermarkov equation
\beq\label{Ermarkov}
\ddot{b}(t)+\omega(t)^2b(t) = \frac{\omega(0)^2}{b(t)^3},\ \text{with}\ b(0)=\dot{b}(0)=0\ .
\eeq

Naturally, we wonder whether a similar scaling dynamics still describes the evolution in the presence of decohrence, i.e., with $D\neq 0$. We can assume that this would be partially true in the sense that a scaling factor (which might obey a different equation) appears in the formula and rescales the position $x\mapsto x/b(t)$. However, in the same time the dissipation will mix the states and the solution of the form \eqref{ScalingD=0}  no longer holds.  

We consider a Gaussian state of the form
\beq\label{GaussianState}
\rho_S(x,y;0) = \mathcal{N}_0 \exp\left[-\frac{1}{2}\l( ax^2+ay^2-2cxy \right)\r]\ ,
\eeq
where $a>c$ are two real numbers and where the normalization factor $ \mathcal{N}_0 = \sqrt{\frac{a-c}{\pi}}$ so that $\text{tr}\l[ \rho_S(0) \r]=\intf dx\ \rho_S(x,x;0) = 1$. For following computations, we rewrite the Gaussian state by introducing a square invertible matrix
\beq\label{GaussianState2}
\rho_S(x,y;0) =  \mathcal{N}_0 \exp\l(-\frac{1}{2}\la X,M_0 X  \ra\r)\ ,
\eeq
where the matrix $M_0$ and the vector $X$ read
$$
M_0=\bmat a & -c \\ -c & a \emat \ , \ X=\bmat x \\ y \emat \ . 
$$
The purity of the Gaussian state is given by the two-dimensional Gaussian integral of equation \eqref{GaussianState2} and reads
\beq\label{PurityGS}
p_0\equiv \text{tr}\l[\rho_S(0)^2\r] = \intf dx\intf dy\ \rho_S(x,y;0)\rho_S(y,x;0) = \sqrt{\frac{a-c}{a+c}}\ .
\eeq

\textbf{Remark.} Here we consider the symmetric complex-valued product $\la X ,Y \ra = X^T Y = x_1y_1 + x_2 y_2 = \la Y ,X \ra$ which it is {\it not} the inner product on $\mathbb{C}^2$ defined as the sesquilinear form $\big(X,Y\big)\equiv X^\dagger Y = x_1^\ast y_1 + x_2^\ast y_2$. However both can be related as follows $\la X ,Y \ra = \big(X^\ast,Y \big)=\l(X^\ast\r)^\dagger Y =  X^T Y$. 

Equation \eqref{rhoSol} gives the density matrix at time $t>0$. 
To compute explicitly the double integral we use the multi-dimensional Gaussian integral method. 
We first rewrite the phase of the integrand as a quadratic form 
\beq\label{phi}
\phi\l(x,y;x_0,y_0;t\r) = \varphi(X) - \frac{1}{2}\la X_0,Q X_0 \ra + \la V,X_0\ra \ ,
\eeq
where $X=\bmat x\\ y \emat$ and $X_0=\bmat x_0\\ y_0 \emat$, and 
\begin{subequations}\label{MultiDim}
\beq\label{varphi}
\varphi(X) = \frac{im\dot{G}_2}{2\hbar G_2}\l(x^2-y^2\r)-D\beta \l(x-y\r)^2 = -\frac{1}{2}\la X,Q_0 X \ra\ ,
\eeq
\beq \label{Qo}
Q_{0}=\bmat \frac{-im \dot{G}_2}{\hbar G_2} + 2D\beta  & -2D\beta  \\ -2D\beta & \frac{+im \dot{G}_2}{\hbar G_2} + 2D\beta  \emat \ , 
\eeq
\beq\label{Q}
 Q = \bmat \frac{-im G_1}{\hbar G_2} + 2D\alpha  & -2D\alpha  \\ -2D\alpha & \frac{+im G_1}{\hbar G_2} + 2D\alpha  \emat + M_0\ ,
\eeq
\beq\label{V} 
V= \bmat -\frac{im (2-\epsilon+\epsilon\dot{G}_2) x}{2\hbar G_2}-2\eta(x-y)-\epsilon\frac{im (1-\dot{G}_2) y}{2\hbar G_2} \\ +\frac{im (2-\epsilon+\epsilon\dot{G}_2) y}{2\hbar G_2}+2\eta(x-y)+\epsilon\frac{im (1-\dot{G}_2) x}{2\hbar G_2}\emat = P\cdot X  ,
\eeq
\end{subequations}
with $\epsilon = \left\{ \begin{array}{ll} 0 \ \text{for}\ \gamma=0 \\  1 \ \text{for}\ \gamma\neq 0 \end{array} \right.$.
We rewrite the vector $V$ using the transfer matrix $P$ 
\beq\label{P}
P = \bmat   -\frac{im (2-\epsilon+\epsilon\dot{G}_2) x}{2\hbar G_2} -2\eta & 2\eta-\epsilon\frac{im (1-\dot{G}_2) }{2\hbar G_2} \\ 2\eta +\epsilon\frac{im (1-\dot{G}_2) y}{2\hbar G_2}& +\frac{im (2-\epsilon+\epsilon\dot{G}_2) x}{2\hbar G_2} -2\eta \emat\ .
\eeq
Notice that for $a=0=c$ equation \eqref{phi} gives the phase of the Feynman-Vernon influence functional \eqref{J}. 
Equations \eqref{rhoSol} and \eqref{phi} leads to the multi-dimensional Gaussian integral form
\begin{align}\label{rhoCompute}
\rho_S(x,y;t)&=\frac{m}{2\hbar\pi G_2}\sqrt{\frac{a-c}{\pi}}e^{-\varphi(X)}\intf dx_0\intf dy_0 e^{-\frac{1}{2}\la X_0,Q X_0 \ra + \la V,X_0\ra}\  \\
&= \frac{m}{2\hbar\pi G_2}\sqrt{\frac{a-c}{\pi}}e^{-\varphi(X)} \frac{\pi}{\sqrt{\det{(Q)}}} e^{\frac{1}{2}\la V,Q^{-1} V \ra}\ ,
\end{align}
where 
\begin{subequations}
\beq
\det{(Q)} = \frac{m^2}{\hbar^2 G_2^2}\l[ G_1^2+\omega_0^2 G_2^2\l( 1+\frac{4\hbar D\alpha}{m\omega_0} \r)  \r]\ ,
\eeq
\beq
Q^{-1} = \frac{1}{\det{(Q)} }\bmat \frac{+im G_1}{\hbar G_2} + 2D\alpha +a & +2D\alpha +c \\ +2D\alpha +c & \frac{-im G_1}{\hbar G_2} + 2D\alpha +a \emat \ .
\eeq
\end{subequations}
Using the expressions of the vector $V$ and the phase $\phi(X)$ in equations \eqref{V} and \eqref{phi}, respectively, we find a more elegant form
\beq\label{rhoFinal}
\rho_S(x,y;t) = \frac{1}{b(t)}\sqrt{\frac{a-c}{\pi}}e^{-\la X , \l(Q_0 + P^T Q^{-1} P \r) X \ra /2}\ .
\eeq
Here, the scaling factor has the form
\beq\label{ScalingFactor}
b(t) = \sqrt{ G_1(t)^2 + (a^2-c^2) \frac{\hbar^2}{m^2}G_2(t)^2\l( 1+ \frac{4}{a+c}D\alpha(t) \r) }\ .
\eeq
and rules the time-evolution of the variance of the position, given that
\begin{equation}\label{Var}
\Delta x(t)^2 = \intf dx\ x^2\rho_S(x,x;t) = \frac{1}{a-c}b(t)^2\ .
\end{equation}

\subsection{Closed-form expression of the survival probability} 

Next, we derive a closed form expression of  the relative purity $S(t)$ in Eq. (\ref{S(t)}), which is identical to the survival probability when the initial state is pure. 
Up to the purity of the initial state,  the unnormalized relative purity $\tilde{S}(t)$ can be computed as follows
\beq
\tilde{S}(t)\equiv \text{tr}\l[ \rho_S(0)\rho_S(t) \r] = \intf dx \intf dy \ \rho_S(x,y;0)\ \rho_S(y,x;t)\ .
\eeq
As we have already computed the expression of the kernel of the density matrix at time $t$, see equation \eqref{rhoSol}, we can compute this double integral using multi-dimensional Gaussian integral technique
\beq
\tilde{S}(t) = \intf dx \intf dy\ \mathcal{N}(t) e^{-\la X , M X  \ra / 2}\ , 
\eeq
with 
\begin{subequations}
\beq
\mathcal{N}(t) = \frac{m}{2\pi\hbar G_2(t)}\frac{a-c}{\pi}\frac{\pi}{\det{(Q)}}\ ,
\eeq
\beq\label{M}
M = Q_0 + P^T Q^{-1} P + M_0\ .
\eeq
\end{subequations}
Gathering the equations above we find
\beq\label{Stilde}
\tilde{S}(t) = \frac{m}{2\hbar G_2(t)}\frac{a-c}{\sqrt{\det{(Q)}}\sqrt{\det{(M)}}}\ .
\eeq

The purity of the Gaussian state \eqref{GaussianState} is $\sqrt{(a-c)/(a+c)}$, leading to the following compact expression for the normalized relative purity
\beq\label{Snorm}
S(t) = \frac{m}{2\hbar G_2(t)}\frac{\sqrt{a^2-c^2}}{\sqrt{\det{(Q)}}\sqrt{\det{(M)}}}\ .
\eeq

\section{Long-time Quantum Decay under the Quantum Brownian motion}

\subsection{General results}

In this section we consider the free expansion of an initial Gaussian state \eqref{GaussianState}. We obtain this case by taking $G_1(t) = 1$ and $G_2(t) = (1-e^{-2\gamma t})/(2\gamma)$, as explained later. After computing the explicit form of relative purity as analogue of the survival probability \eqref{Stilde} we can demonstrate the following asymptotic forms for two distinct regimes
\begin{subequations}\label{S(t)Asympt}
\beq\label{S(t)Asympt1}
\tilde{S}(t) \approx \sqrt{3}\frac{m}{D\hbar t^2} = \frac{\tau_\beta}{\tau_R}\sqrt{\frac{\tau_R}{t}}\ , \ \text{when}\ \tau_D\ll t\ll \gamma^{-1}\ ,
\eeq
\beq\label{S(t)Asympt2}
\tilde{S}(t) \approx \frac{\hbar\gamma}{2k_B T}\frac{1}{\sqrt{\gamma t}}=\frac{\tau_\beta}{\tau_R}\sqrt{\frac{\tau_R}{t}} \ , \ \text{when}\  t\gg \gamma^{-1}\ ,
\eeq
\end{subequations}
which confirm the asymptotic formulae given in the main body of the paper. In equation \eqref{S(t)Asympt} we introduced the decoherence time $\tau_D$ that will be discuss in next. We also introduced two time scales $\tau_\beta\equiv \hbar/(k_B T)$ and $\tau_R=\gamma^{-1}$ that characterize the relaxation of the system and baths, respectively. The details of calculations will be shown hereafter.

Notice that these results can be easily generalized for displaced Gaussian states
\beq\label{GaussianStateDisplaced}
\rho_S(x,y;0) = \mathcal{N}_0 \exp{\l[-\frac{1}{2}\l( a(x-b)^2+a(y-b)^2-2c(x-b)(y-b) \r)\r]}\ ,
\eeq
corresponding to a general coherent squeezed thermal state. It suffices to modify the equations \eqref{MultiDim} to include the linear terms in the phase of the state \eqref{GaussianStateDisplaced}. Following the same multi-dimensional Gaussian integral technique, we find the same asymptotic equations \eqref{S(t)Asympt}. As a result, the method also applies for Schr\"odinger cat states of the form
\beqa
\psi_0(x) = \mathcal{N}_\sigma \l( e^{-\frac{(x-r)^2}{2\sigma^2}} + e^{-\frac{(x+r)^2}{2\sigma^2}} \r)\ ,
\eeqa
for which  the asymptotic decay is set by equations \eqref{S(t)Asympt}. The only difference here is the value of the decoherence time $\tau_D$ which depends on the spreading of the initial state. This case is an important illustration as it extends the well-known results in \cite{Zurek91} to long-time asymptotic regimes for all value of distance $r$,  that sets  the size of the Schr\"odinger cat state. 
 
\subsection{Decoherence time} 

To understand more deeply the long-time asymptotic, we first discuss the short-time expansion of the survival probability. We  first consider an initial pure state by taking $c=0$. Taking $L=\hat{x}$ in \eqref{S(t)order1}, with corresponding \( \gamma_x=2D=4m \gamma k_BT/\hbar^2= 2\gamma/\lambda_\beta^2 \), we find that the decoherence time is given by
\beq\label{tauDpurestate}
\tau_D = \frac{\lambda_\beta^2}{2\gamma\Delta x^2}\ ,
\eeq
where $\Delta x^2 = a^{-1}$ is the width of the initial state and $\lambda_\beta= \hbar/\sqrt{2m k_B T}$ is the de Broglie length, and 
$$
\Delta x^2 = \text{tr}\l[\hat{x}^2\rho_S(0)\r]-\text{tr}\l[\hat{x}\rho_S(0)\r]^2\ . 
$$
is the variance of the Lindblad operator $L=\hat{x}$ with respect to the initial state. To show the more general right hand side of \eqref{tauDGeneral}, we used a universal formula derived in \cite{CBCdC16} valid for dephasing Markovian dynamics. 
For a mixed state with purity smaller than one (see e.g., equation \eqref{GaussianState} with $c\neq 0$), we find the following formula
\beq\label{tauDGeneral}
\tau_D = \frac{\lambda_\beta^2}{2\gamma\widetilde{\Delta} x^2} \text{tr}\l[\rho_S(0)^2\r] \ ,
\eeq  
where 
$$
\widetilde{\Delta} x^2 =  \text{tr}\l[\hat{x}^2\rho_S(0)^2\r] - \text{tr}\l[\l(\hat{x}\rho_S(0)\r)^2\r] = \frac{1}{2}\intf dx\intf dy\ (x-y)^2\rho_S(x,y;0)^2\ .
$$
More specifically, for the Gaussian state \eqref{GaussianState} we find
\beq\label{DeltaxGeneralGaussian}
\widetilde{\Delta} x^2 = \frac{1}{2(a+c)}\sqrt{\frac{a-c}{a+c}}\ ,
\eeq
which is equal to the squared variance $\Delta x^2 = 1/(2a)$ if the state is pure, i.e., if $c=0$.

There is a caveat to bear in mind at this point. As is well known, the Caldeira--Leggett master equation is not of Lindblad form. Nonetheless, the addition of a minimally invasive term, \( - \gamma\left[\hat{p},\left[\hat{p},\rho_S(t)\right]\right]/8m k_B T= - (\gamma \lambda_\beta^2/4\hbar^2)\left[\hat{p},\left[\hat{p},\rho_S(t)\right]\right] \), allows the master equation to be written in Lindblad form with a single damping constant \(\gamma\) and Lindblad operator
\[\hat{A}= \sqrt{2} \frac{\hat{x}}{\lambda_\beta}+ i\frac{\lambda_\beta\hat{p}}{\hbar \sqrt{2}}\,.\]
Since the microscopic derivation of the Caldeira--Leggett master equation holds under the assumption of large temperature, i.e. that the thermal time \( \hbar/k_BT \ll \tau_S\), where \( \tau_S \) is a typical time for the system, the term being added is typically small, and similarly, it can be argued that the last term of  \( \hat{A} \) is also negligible. Namely, typically \( \hat{p}\sim m \lambda_\beta/\tau_S \), whence the second term is of order \( \tau_\beta/2\tau_S \). Alternatively, observe that our definition of \( \tau_D \) applied to the dissipator with \( \hat{A} \) gives
\begin{equation}
\tau_D^{-1}= \frac{\gamma}{ \mathrm{tr}[ \rho_S(0)^2]} \left\{ \frac{2 \widetilde{\Delta}x^2}{ \lambda_\beta^2} + \frac{ \lambda_\beta^2}{2\hbar^2} \widetilde{\Delta}p^2+ \mathrm{tr}\left[ \rho_S(0)^2\right]\right\}\,.
\end{equation}
Under the conditions of validity of the Caldeira--Leggett master equation for quantum Brownian motion the first term dominates, thus reducing to Eq. (\ref{tauDGeneral}).
\subsection{Quantum decay  of a Gaussian state in a free expansion under quantum Brownian dynamics}

We consider the free expansion at time equal zero, i.e., $\omega(t)=0,\ t\geq 0$. Let  us first determine the fundamental solutions of equation \eqref{ClassEq} and the functions in equation \eqref{functionsJ}
\begin{subequations}
\beq
G_1(t)=1,\ G_2(t)=\frac{1}{2\gamma}\left(1-e^{-2\gamma t}\right),\ \dot{G}_2(t)=e^{-2\gamma t}\ ,
\eeq
\beq
\alpha(t)=\frac{4\gamma t-3+4 e^{-2\gamma t}-e^{-4\gamma t}}{4\gamma\l(1-e^{-2\gamma t}\r)^2}\ ,
\eeq
\beq
\beta(t) = \frac{1- 4 e^{-2\gamma t}+\l(4\gamma t+3\r)e^{-4\gamma t}}{4\gamma\l(1-e^{-2\gamma t}\r)^2}\ ,
\eeq
\beq
\eta(t) = \frac{1- 4 e^{-2\gamma t}-4\gamma t e^{-4\gamma t}}{4\gamma\l(1-e^{-2\gamma t}\r)^2}\ .
\eeq
\end{subequations}
For short-time, we have to expand these functions at first order in the regime $\gamma t\ll 1$:
\begin{subequations}\label{AsymptFct}
\beq
G_1(t)=1,\ G_2(t)\approx t,\ \dot{G}_2(t)\approx 1-2\gamma t\ ,
\eeq
\beq
\alpha(t)\approx \frac{t}{3}\ ,\ \beta(t) \approx \frac{t}{3}\ ,\ \eta(t) \approx\frac{t}{6}\ .
\eeq
\end{subequations}
and for long-time, we expand for $\gamma t\gg 1$
\begin{subequations}
\beq
G_1(t)=1,\ G_2(t)\approx \frac{1}{2\gamma},\ \dot{G}_2(t)\approx 0\ ,
\eeq
\beq
\alpha(t)\approx t\ ,\ \beta(t) \approx \frac{1}{4\gamma}\ ,\ \eta(t) \approx \frac{1}{4\gamma}\ .
\eeq
\end{subequations}

First, we consider the undamped limit taking $\gamma=0$ in \eqref{MasterEq2} (pure decoherence). Equation \eqref{AsymptFct} leads to the asymptotic form \beq\label{SfreeSTUndamped}
S_{\text{free}}(t)\approx 1-\frac{t}{\tau_D}\ ,\ \gamma t\gg \tau_D \ ,
\eeq 
and
\beq\label{SfreeLTUndamped}
S_{\text{free}}(t)\approx \frac{\sqrt{3}\tau_D\tau_S}{t^2}\ ,\ \gamma t\ll \tau_D\ , 
\eeq 
where 
\beq\label{tauDS}
\tau_D=\frac{(a+c)}{D}\ ,\ \tau_S = \frac{m}{\hbar\sqrt{a^2-c^2}}\ ,
\eeq
and $\tau_S$ is the characteristic time of the unitary evolution of a Gaussian wave packet for which the scaling factor is 
$$
b_0(t)=\sqrt{1+\frac{t^2}{\tau_S^2}}\ .
$$
In the presence of pure decoherence (without dissipation), Equation \eqref{ScalingFactor} now reads
$$
b(t)=\sqrt{1+\frac{t^2}{\tau_0^2}\l(1+\frac{4t}{3\tau_D}\r)}\ .
$$
For non-zero $\gamma$ satisfying $\gamma (a+c)\ll D$  equation \eqref{SfreeLTUndamped} determines the asymptotic form of the relative purity $S(t)$ for $t$ not too large
$$
\gamma t\ll 1 \ll \frac{D t}{a+c} \Rightarrow \frac{a+c}{D}\ll t\ll \frac{1}{\gamma}\ .
$$

Considering now damping dynamics, we have two other regimes where the relative purity behaves like
\beq\label{SfreeST}
S_{\text{free}}(t)\approx 1-\l(\frac{D}{(a+c)\gamma}-1\r)\gamma t\ ,\ \text{for}\ \gamma t\ll 1 \ ,
\eeq 
and
\beq\label{SfreeLT1}
S_{\text{free}}(t)\approx \sqrt{\frac{a+c}{a-c}}\sqrt{\frac{4m^2\gamma^4}{\hbar^2 D\l(D+(a+c)\gamma\r)}}\times\frac{1}{\sqrt{\gamma t}}\ ,\ \text{for}\ \gamma t\gg 1\ . 
\eeq 

The decoherence time appears in Equation \eqref{SfreeST} 
\beq\label{tauD}
\tau_D = \frac{(a+c)}{D-(a+c)\gamma}\approx \frac{(a+c)}{D}\ ,
\eeq
is positive as in the large temperature regime the bath correlation $\tau_T=\frac{\hbar}{2\pi k_B T}$ is large compared with the characteristic evolution time of the system $\tau_S$ defined in \eqref{tauDS}, i.e., 
$\tau_S \ll \tau_D \Rightarrow D\gg (a+c)\gamma $. This justifies the last approximation in \eqref{tauD}. \\
In the same fashion, we can give the asymptotic of equation \eqref{SfreeLT1} 
\beq\label{SfreeLT2}
S_{\text{free}}(t)\approx \sqrt{\frac{a+c}{a-c}}\frac{2m\gamma^2}{\hbar D}\frac{1}{\sqrt{\gamma t}}=\sqrt{\frac{a+c}{a-c}}\ \frac{\hbar\gamma}{k_B T}\frac{1}{\sqrt{\gamma t}}\ ,\ \text{for}\ \gamma t\gg 1\ ,
\eeq
where the amplitude is proportional to the inverse of the initial purity and to the coefficient $\hbar\gamma/(k_B T)$ which is assumed to be small in large temperature limit. 
 
%\begin{figure}
%\includegraphics[width= 0.5\columnwidth]{Fig_Sfree.pdf}\ \ \ \\
%\caption{{\bf Initial Gaussian state free expansion survival probability. } 
 %\label{Fig1}}
%\end{figure}

It is also interesting to look at the density profile and at the off-diagonal density elements for long-time. 
It is easy to show that for $t\gg \gamma^{-1}$ we obtain the standard classical Brownian diffusion density profile
\beq\label{DensityProfileLongTime}
\rho_S(x,x;t)\approx \frac{1}{\sqrt{2\pi \sigma t}} \exp\l(-\frac{x^2}{2\sigma t}\r)\ ,
\eeq
where the diffusion coefficient $\sigma = \frac{\hbar^2 D}{2m^2\gamma^2} = \frac{k_B T}{m\gamma}$ is proportional to the mobility $\mu=\frac{1}{m\gamma}$. Consistently, the asymptotic of the width of the wave packet \eqref{Var} gives $\sqrt{\sigma t}$ as expected \cite{Weiss}. 

Now we look at the asymptotic of the off-diagonal density matrix. As expected, we find Gaussian correlations with a width proportional to the temperature
\beq \label{OffDiagLongTime}
\rho_S(x,y;t)\approx \frac{1}{\sqrt{2\pi \sigma t}} \exp\l[-\frac{(x-y)^2}{2\lambda_{dB}^2}\r]\ ,\ \text{with}\ x\neq y\ , 
\eeq 
where $\lambda_{dB} = \sqrt{\frac{\hbar^2}{2mk_B T}}$ is the de Broglie thermal wavelength.   \\

\subsection{Applications} 

The method developed above allows to treat some particular quantum states  of prominent relevance in physics. The Gaussian state introduced in equation \eqref{GaussianState2} is particularly useful as for different choices of parameters $a$ and $c$ it covers the following cases
\begin{itemize}
\item Taking $a=\xi_0^{-2}$ and $c=0$ with $\xi_0 = \sqrt{\frac{\hbar}{m\omega_0}}$, \eqref{GaussianState} is the ground state of the quantum harmonic oscillator (QHO), 
\beq
\rho_S(x,y;0) = \sqrt{\frac{m\omega_0}{\pi\hbar}}e^{-\frac{m\omega_0}{2\hbar}\l(x^2+y^2\r)}\ .
\eeq 
\item We can get the Mehler kernel by choosing $a=\xi_0^{-2}(1+u^2)/(1-u^2)$ and $c=\xi_0^{-2} 2u/(1-u^2)$
\beq\label{MehlerKernel}
K(x,y;u)=\sqrt{\frac{m\omega_0}{\pi\hbar(1-u^2)}} e^{-\frac{m\omega_0}{2\hbar}\l[ \frac{1+u^2}{1-u^2}(x^2+y^2)-\frac{2 u}{1-u^2} x y \r]}\ .
\eeq
This state is not normalized $\text{tr}\l(\rho_0\r) = |1-u|^{-1}$ so all the solutions given by equations \eqref{rhoFinal} and \eqref{Snorm} should be multiplied by this factor $|1-u|^{-1}$. 
\item Using the Mehler formula 
\beq\label{MehlerFormula}
K(x,y;u) =\sum_{n=0}^{\infty}u^n \phi_n(x)\phi_n(y) \Rightarrow    \phi_n(x)\phi_n(y) = \frac{1}{n!}\frac{d^n}{du^n}K(x,y;u) \Big|_{u=0} \ ,
\eeq
we can tackle not only the ground-state but an arbitrary excited state $|\psi_n\ra$ of the QHO $\rho_0 = |\psi_n\ra\la\psi_n|$, where 
$$
\la x|\psi_n\ra = \psi_n(x)=\frac{1}{\l(2^n n!\r)^{1/2}}\l(\frac{m\omega_0}{\pi\hbar}\r)^{1/4}H_n\l(\sqrt{\frac{m\omega_0}{\hbar}}x\r)e^{-\frac{m\omega_0}{2\hbar}x^2}
$$ 
are the normalized Hermite functions. 
\item In a similar fashion, by taking $ u = e^{-\beta\hbar\omega_0}$ where $\beta=1/(k_B T)$ the Mehler formula \eqref{MehlerFormula} (multiplied by $\sqrt{u}=e^{-\beta\hbar\omega_0/2}$ where $\hbar\omega_0/2$ is the ground state energy of the QHO) gives the kernel of the thermal state of a quantum particle in the canonical ensemble  
$$
\la x|e^{-\beta H}|y\ra = e^{-\beta\hbar\omega_0/2} \times K(x,y;u=e^{-\beta\hbar\omega}) = \sum_{n=0}^{\infty} e^{-\beta \epsilon_n}\phi_n(x)\phi_n(y)\ ,
$$
that upon  normalization provides the density matrix of the canonical thermal state
\beq
\rho_S(x,y;\beta) = \frac{1}{Z}\la x|e^{-\beta H}|y\ra = \sqrt{\frac{m}{2\pi\hbar\sinh(\beta\hbar\omega_0)}}e^{-\frac{m\omega_0}{2\hbar}\l( \coth(\beta\hbar\omega_0)(x^2+y^2)-\cosh(\beta\hbar\omega_0)xy \r)}\ ,
\eeq
where the partition function $Z=(1-e^{-\beta\hbar\omega_0})^{-1}$.
\item A stationary state can be described by \eqref{GaussianState} with $a = \frac{\sigma_p^2}{\hbar^2}+\frac{1}{4\sigma_x^2}$ and $c= \frac{\sigma_p^2}{\hbar^2}-\frac{1}{4\sigma_x^2}$ and the density matrix
\beq
\rho_S(x,y;0) = \frac{1}{\sqrt{2\pi\sigma_x^2}}e^{-\l[ \frac{1}{8\sigma_x^2}\l(x+y\r)^2 +\frac{\sigma_p^2}{\hbar^2} (x-y)^2 \r]}\ ,
\eeq
where the variance of the position and momentum are respectively given by $\sigma_x^2 = k_B T/(m\omega_0^2)$ and $\sigma_p^2 = m k_B T $. 
\end{itemize}

\subsection{Long-time asymptotic behavior: general argument}

Consider $\rho_S(x,y;0)$ to be a general Gaussian wave packet with initial spreading $\Delta x(0) \equiv \Delta x$. The method presented below relies on two assumptions. First the characteristic time scale of the system $\tau_c\equiv m\Delta x^2/\hbar$ is large compared to the one of the thermal bath  $\tau_\beta = \hbar/(k_B T)$ ,or equivalently $\lambda_\beta \ll \Delta x$. This means that the spatial distribution of the original state is macroscopic while the de Broglie length is microscopic (high temperature limit). The second assumption states that the relaxation time of the environment $\tau_R \equiv \gamma^{-1}$ is small compared to $\tau_\beta$ consistently with the high temperature limit. This implies that the dynamics is separated into three distinct regimes, where $\tau$ is the characteristic time  
\begin{enumerate}
\item For $  \tau \ll \tau_D $, thee particle is in the Zeno regime and the survival probability expansion behaves as $S(t)\approx 1-t/\tau_D$. 
\item The first long-time regime holds when $ \tau_D \ll \tau \ll \tau_R$. In this case, the system is undamped and experiences decoherence. The off-diagonal density matrix decay exponentially in time \cite{Zurek91} 
\beq\label{rhoLT1}
\rho_S(x,y;t)\approx \rho_S(0,0;t)\times \exp{\l[-D t (x-y)^2\r]} \exp{\l[i\phi(x,y)\r]}\ ,
\eeq
where $\phi$ is the complex part of the phase. In this case the normalization factor of the density matrix scales as
\beq\label{DeltaxLT1}
\Delta x(t) \approx \sqrt{\frac{4Dt^3}{3}}\ .
\eeq
\item As the time becomes large compared to the relaxation time of the environment $\tau\gg \tau_R$, the system is largely damped and the off-diagonal density matrix converges to a stationary solution
 \beq\label{rhoLT2}
\rho_S(x,y;t)\approx \rho_S(0,0;t)\times \rho_S(0,0;t) \exp{\l[-\frac{(x-y)^2}{2\lambda_\beta^2}\r]} \exp{\l[i\phi(x,y)\r]}\ ,
\eeq
and the normalization factor is the diffusive 
\beq\label{DeltaxLT2}
\Delta x(t) \approx \sqrt{\frac{D\hbar^2 }{m\gamma^2}t} =\sqrt{\frac{2k_B T}{m\gamma} t}\ .
\eeq
\end{enumerate}   

From the integral representation of the unormalized survival probability $\tilde{S}(t)$, we find the following decomposition
\begin{align}\label{S(t)Decomp}
\tilde{S}(t) &= \intf dx\intf dy\ \rho_S(x,y;0)\rho_S(x,y;t) \\
&= \intf dx \l\{ \int_{-\infty}^{-\Delta x}dy+\int_{\Delta x}^{+\infty}dy + \int_{-\Delta x}^{+\Delta x}dy \r\} \rho_S(x,y;0)\rho_S(x,y;t) \ ,
\end{align}
where the initial variance $\Delta x$ sets the scales of the screening of the density distribution. It is clear that the first two integral tends to zero in the long-time regimes as
\begin{align*}
 \exp{\l[-D t (x-y)^2\r]} = \exp{\l(-t/\tau_D\r)} & \rightarrow 0,\ \text{when}\ t\gg \tau_D = \frac{1}{D\Delta x}\ , \\
 \exp{\l[-\frac{(x-y)^2}{2\lambda_\beta^2}\r]}& \rightarrow 0,\ \text{as}\ \lambda_\beta \ll \Delta x\ .
\end{align*}
Notice also that 
$$
\frac{\Delta x^2}{\lambda_\beta^2} = \frac{m\gamma k_B T \Delta x^2}{\hbar^2}\frac{1}{\gamma}=\frac{D \Delta x^2}{\gamma}=\frac{\tau_R}{\tau_D}\ll 1\ ,
$$ 
consistently with the previous assumptions. 

Now, looking at the third integral in \eqref{S(t)Decomp}, after regularizing the integral by multiplying the Gaussian term in equation \eqref{rhoLT1} by the factor $\sqrt{4\pi/Dt}$ and by $\sqrt{2\pi \lambda_\beta^2}$  in equation  \eqref{rhoLT2}, we approximate the integrand by a delta-distribution and obtain 
 \begin{subequations}\label{S(t)Approx1}
\beq
\tilde{S}(t) \approx \sqrt{\frac{4\pi}{Dt}}\intf dx\ \rho_S(x,x;0)\rho_S(x,x;t),\ \text{when}\ \tau_D\ll t\ll\tau_R\ , 
\eeq
\beq
\tilde{S}(t) \approx \sqrt{2\pi \lambda_\beta^2}\intf dx\ \rho_S(x,x;0)\rho_S(x,x;t),\ \text{when}\ \tau_R\ll t\ .
\eeq
\end{subequations}
These equations mean that at long times of evolution the coherence term vanishes and the integral reduces to the overlap between the initial and final density profiles. The long-time density profile is determined by the Hamiltonian dynamics with the modified scaling factor 
\beq
\rho_S(x,x;t) \approx \frac{1}{\sqrt{2\pi \Delta x(t)^2}} \exp{\l(-\frac{x^2}{2\Delta x(t)^2}\r)}\ ,
\eeq
where the asymptotic of the variance are given by equations \eqref{DeltaxLT1} and \eqref{DeltaxLT2} and we clearly have $\Delta x(t)\gg \Delta x(0)$. Therefore the initial density profile can be approximate as a delta-distribution in equation \eqref{S(t)Approx1}, leading to equation \eqref{S(t)Asympt}. 
%These results confirm the previous results obtained for a the Gaussian wavepacket \eqref{GaussianStateGen}. Another important remark concerns the universality of the asymptotic given in equations \eqref{S(t)Approx2}. Indeed, the coefficients do not depend on the initial variance of the wavepacket. Hence, this universal feature holds for any thermal coherent squeezed state. This could be be verify in a experimental platform []. Applications are also straightforward, the value of the slops obtained (see Fig 1 in the main body of the paper) gives information on the two time scales $\tau_\beta$ and $\tau_\gamma$ and consequently on the values of the temperature $T$ and of the damping constant $\gamma$.  

\begin{figure}
 \includegraphics[width= 0.3\columnwidth]{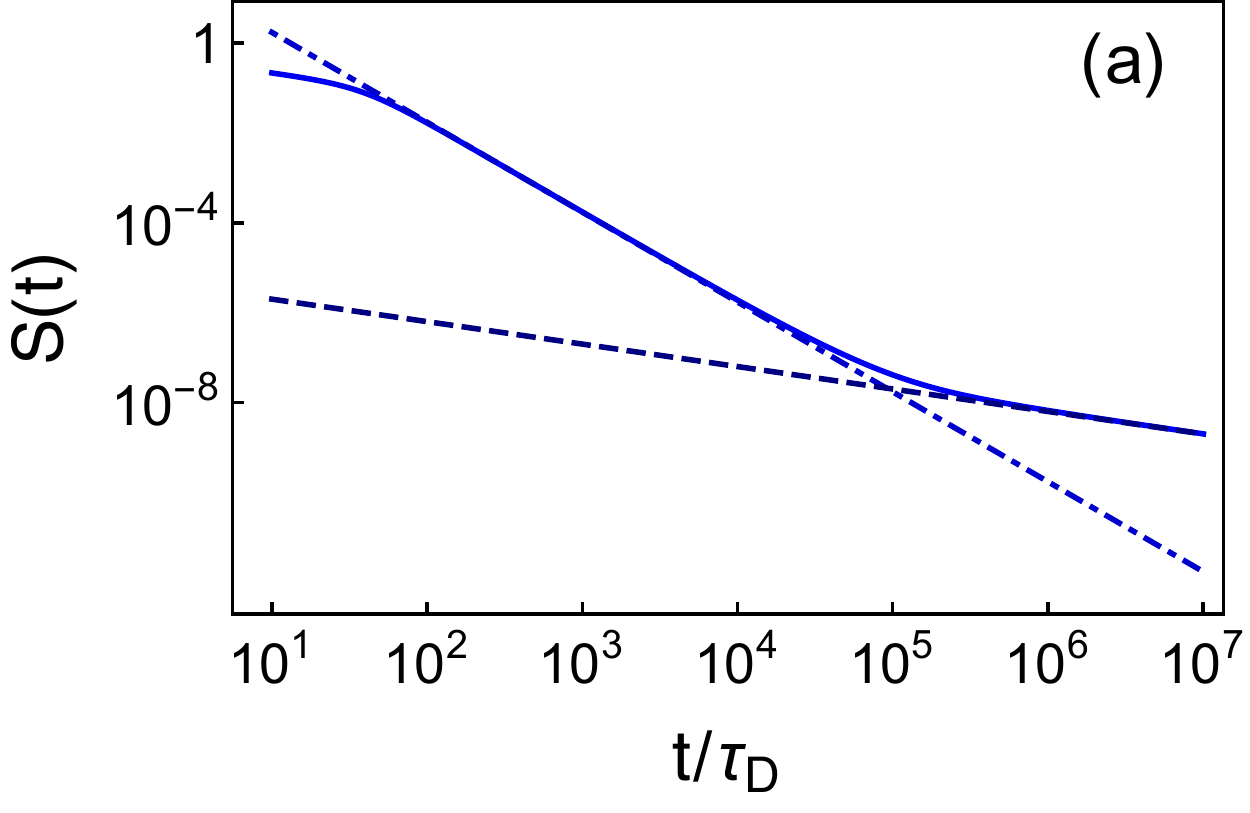} \includegraphics[width= 0.3\columnwidth]{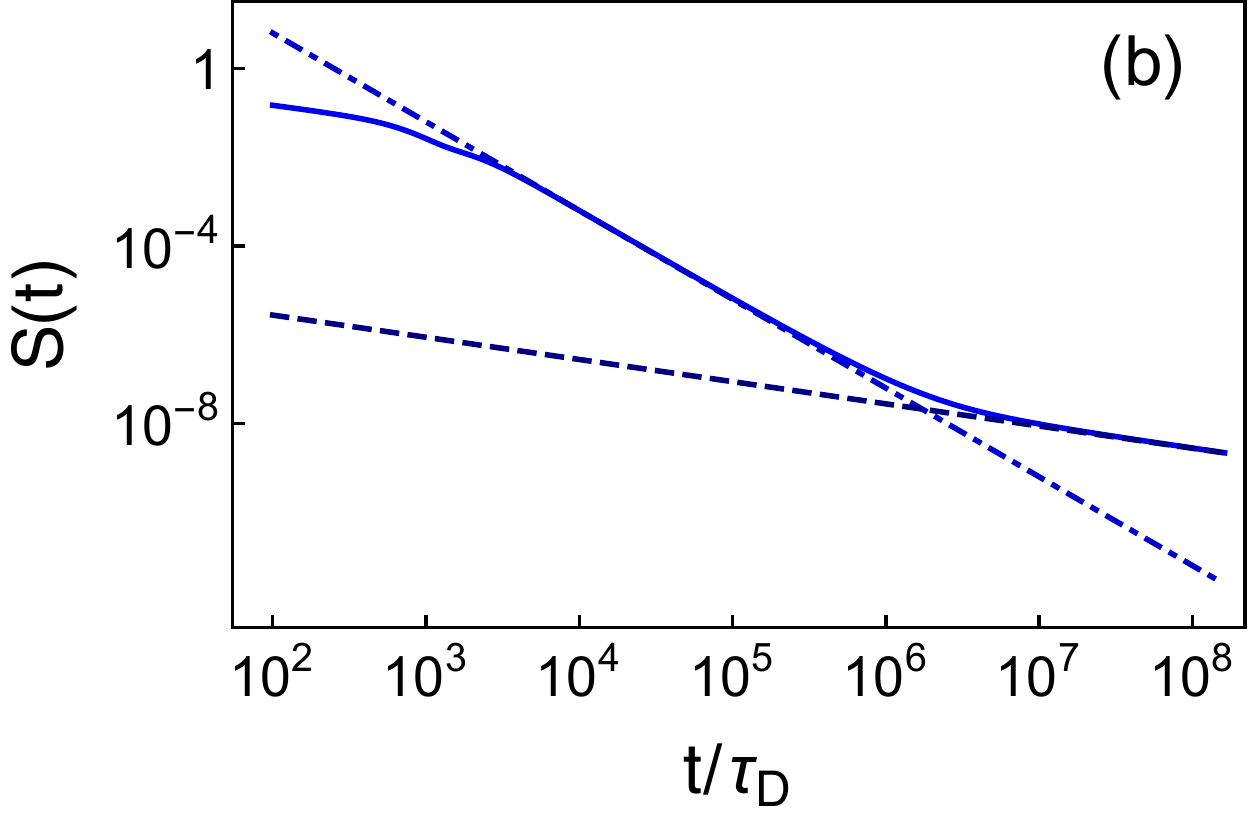} \includegraphics[width= 0.3\columnwidth]{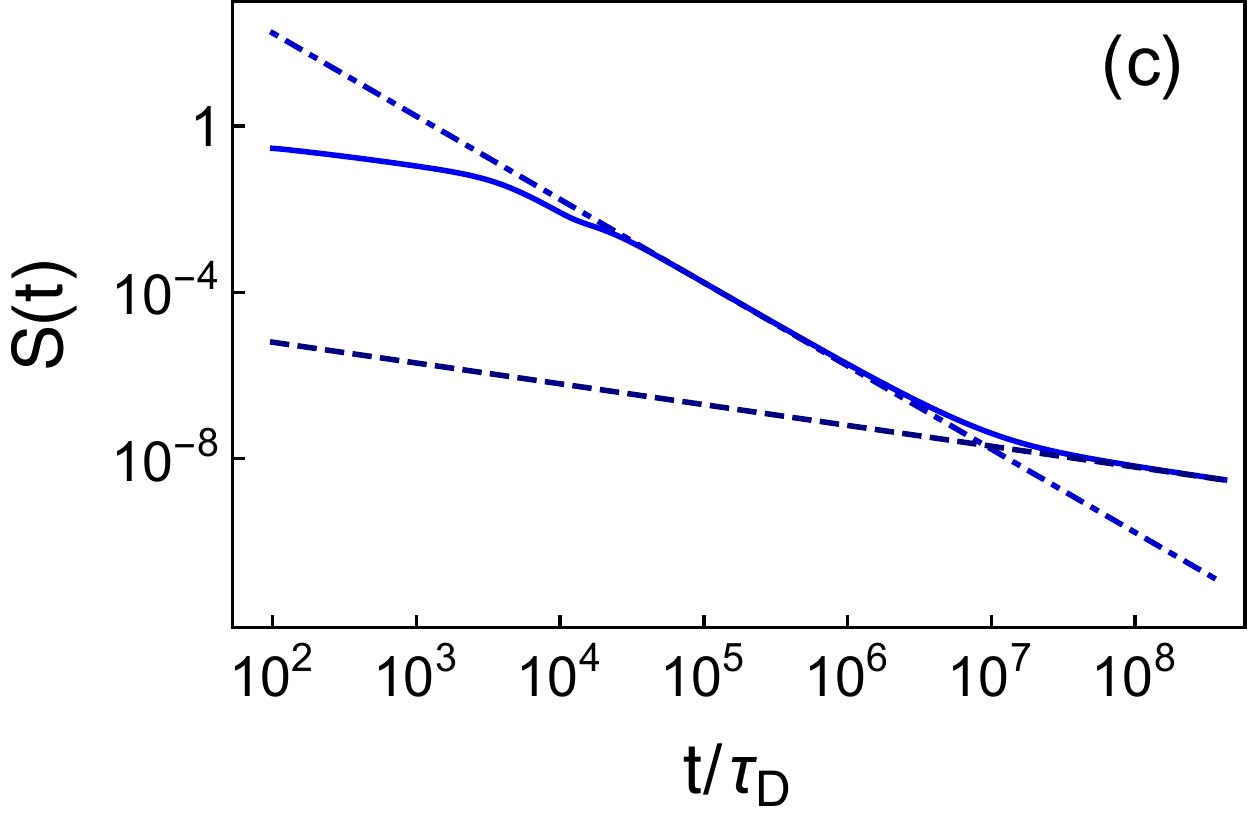}
\caption{{\bf Decay of the survival probability under quantum Brownian motion.}  The survival probability of a Schr\"odinger cat-state (continuous line) is displayed in a log-log scale as a function of time  in units of the decoherence time.  Its behavior is accurately reproduced by the long-time asymptotic forms $t\gg\tau_D$ in the two distinct regimes $\gamma t\ll 1$ (dotted-dashed line) and $\gamma t\gg 1$ (dashed line), as shown for the choice of parameters $\gamma = 10^{-3}$  and $D\sigma^2=100$ with $\hbar/m\sigma^2$ as frequency unit and  (a) $r=0$, (b) $r=3$, and  (c) $r=7$ in units of $\sigma$. }\label{Fig1}
\end{figure}

\subsection{Application to Schr\"odinger cat states}

Consider a pure state $\rho_S(x,y;0)=\psi_0(x)\psi_0(y)^\ast$ made of a superposition of two Gaussian wave packets centered respectively in $x=-r$ and $x=+r$,
\beq\label{TwoGaussianState}
\psi_0(x) = \mathcal{N}_\sigma \l( e^{-\frac{(x-r)^2}{2\sigma^2}} + e^{-\frac{(x+r)^2}{2\sigma^2}} \r)\ ,
\eeq
where the normalization factor is
$$
\mathcal{N}_\sigma = \frac{1}{\sqrt{2\sqrt{\pi\sigma^2}(1+e^{-\frac{r^2}{\sigma^2}})}}\ .
$$
The variance of the position is given by
\beq
\Delta x^2 = \frac{r^2}{1+e^{-\frac{r^2}{\sigma^2}}} + \frac{\sigma^2}{2}\ .
\eeq
From the exact expression for the decoherence time in equation \eqref{tauDpurestate}, we find the leading asymptotic behavior
\beq
\tau_D = \frac{\lambda_\beta^2}{2\gamma\Delta x^2} \approx \frac{\lambda_\beta^2}{2\gamma r^2},\ \text{as}\ r\gg \sigma\ ,
\eeq
which reproduces the Zurek prediction. In the other limit $r\ll \sigma$, we find the prediction we obtained for a single Gaussian wave packet, see equation \eqref{tauD}. 

Using a similar multi-dimensional Gaussian integral approach, we compute the survival probability and show the asymptotics in equation \eqref{S(t)Asympt}, see Fig. \ref{Fig1}. This demonstrates that the general argument developed in the previous section applies for the initial state \eqref{TwoGaussianState}.

\end{widetext}
\end{document}